\documentclass[showpacs,aps,prl,superscriptaddress]{revtex4}

\usepackage{graphicx,amsfonts}
\usepackage{latexsym}

\begin{document}

\title{Interlaced solitons and vortices in coupled DNLS lattices}
\author{J. Cuevas}
\affiliation{Grupo de F{\'i}sica No Lineal, Universidad de
Sevilla. Departamento de F{\'i}sica Aplicada I. Escuela
Universitaria Polit{\'{e}}cnica, C/ Virgen de {\'{A}}frica, 7, E-41011
Sevilla, Spain}

\author{Q.E. Hoq}
\affiliation{Department of Mathematics, Western New England College, Springfield, Massachusetts 01119, USA}

\author{H. Susanto}
\affiliation{School of Mathematical Sciences, University of Nottingham, University Park, Nottingham, NG7 2RD, United Kingdom}
\author{P.G. Kevrekidis}
\affiliation{Department of Mathematics and Statistics, University of Massachusetts, Amherst, Massachusetts 01003-4515, USA}

\begin{abstract}
In the present work, we propose a new set of coherent structures that arise
in nonlinear dynamical lattices with more than one components, namely
interlaced solitons. These are waveforms in which in the relevant
anti-continuum limit, i.e.\ when the sites are uncoupled, one component
has support where the other component does not. We illustrate systematically
how one can combine dynamically stable unary patterns to create
ones such for the binary case of two-components. In the one-dimensional
setting, we provide also a detailed theoretical analysis of the existence
and stability of these waveforms, while in higher dimensions, where such
analytical computations are far more involved, we resort to corresponding
numerical computations. Lastly, we perform direct numerical simulations
to showcase how these structures break up, when exponentially or
oscillatorily unstable, to structures with a smaller number of
participating sites.
\end{abstract}


\maketitle

\section{Introduction}

One of the highly active areas of investigation of
Hamiltonian nonlinear systems over the past decade has been
the examination of nonlinear dynamical lattices of the
discrete nonlinear Schr{\"o}dinger (DNLS) type. Chiefly, this
development has arisen due to the multitude of applications
of pertinent models that have emerged in areas such as
nonlinear optics and atomic physics.

More specifically, in the optical context, the setting of
fabricated AlGaAs waveguide arrays \cite{7} has been one
of the most prototypical ones for the application of DNLS
models. There, the interplay of discreteness and nonlinearity
revealed many interesting features including Peierls-Nabarro potential
barriers, diffraction and diffraction management \cite{7a},
and gap  solitons \cite{7b}, among others;
see also the reviews \cite{review_opt,general_review}
and references therein.

Another recent development, which also promoted the
analysis of discrete systems in connection with nonlinear
optics was the proposal \cite{efrem} and creation \cite{moti1,moti2}
of optically
induced photonic lattices in photorefractive crystals such as SBN.
This paved the way for the observation of a large set of
exciting nonlinear wave related phenomena in such
crystals. As a representative subset, we mention here
the formation of patterns such as
dipole \cite{dip}, quadrupole \cite{quad} and necklace \cite{neck}
solitary waves, impurity modes \cite{fedele}, discrete vortices
\cite{vortex1,vortex2}, rotary waves \cite{rings}, higher order
Bloch modes \cite{neshev2} and gap vortices \cite{motihigher},
two-dimensional (2D) Bloch oscillations and Landau-Zener tunneling
\cite{zener}, wave formation in honeycomb
\cite{honey}, hexagonal \cite{rosberg2} and quasi-crystalline
lattices \cite{motinature1}, and recently the study of
Anderson localization in disordered photonic lattices
\cite{motinature2}. Although this setting is mostly studied
in the continuum context with a periodic potential (and sometimes
in the presence of the inherent crystal anisotropy), it has also spurred
a number of studies in the DNLS context with the saturable
photorefractive nonlinearity \cite{boris1,us1}.

Lastly, another physical realization of such nonlinear dynamical
lattices arose over the past few years in atomic physics through
the examination of Bose-Einstein condensates (BECs) trapped in periodic
potentials. There, once again, a reduction of the relevant model
can be formulated in the tight-binding approximation within the
mean-field limit, reducing the so-called Gross-Pitaevskii equation with a
periodic potential to a
genuinely discrete nonlinear Schr{\"o}dinger equation
\cite{ourbook}.

In both the nonlinear optical and in the atomic physics setting
discussed above, multi-component systems were also examined
in recent investigations. More specifically, the first
observations of discrete vector solitons in optical waveguide
arrays were reported in \cite{christo}, the emergence
of multipole patterns in vector photorefractive crystals
was presented in \cite{zhig_two}, while numerous experiments
with BECs were directed towards studies of mixtures of different
spin states of $^{87}$Rb \cite{myatt,us_dh} or $^{23}$Na \cite{stamper}
and even ones of different atomic species such as
$^{41}$K--$^{87}$Rb \cite{KRb} and $^{7}$Li--$^{133}$Cs
\cite{LiCs}. It should be noted that while the above BEC
experiments did not include the presence of an optical lattice,
the addition of such an external
optical potential is certainly feasible within the present
experimental capabilities \cite{morsch3}.

Our aim in the present work is to propose and analyze a family
of solutions particular to multicomponent (in particular, binary,
although more-component generalizations are certainly possible)
systems of DNLS equations. We dub these proposed solutions
``interlaced'' discrete solitons and vortices, a name stemming
from the feature that the profiles of the modes in the two
interacting components will have a vanishing intersection
of excited sites in the extreme discrete limit of zero
coupling between adjacent nodes of the lattice. In these structures, the
first component will be excited where the second component
is not and vice-versa. In the one-dimensional case, we show
how to interlace in a stable fashion simple, as well as more
elaborate, bound states of the system \cite{pkf1}. For such solutions,
we consider their existence and stability properties also
from an analytical point of view, using as a starting point
the anti-continuum limit (of no-coupling between the sites).
Then we generalize our considerations to higher dimensional
settings, showcasing the potentially stable interlacing of
more elaborate structures, such as discrete vortices \cite{pkf2}
(but also of vortices with non-vortical structures). We present
detailed stability diagrams of such interlaced structures,
and also examine their dynamics when they are found to
be unstable.

Our presentation
is structured as follows. In section II, we present the model
and general mathematical setup. In section III, we illustrate
both analytically and numerically the properties of such structures
in 1d settings. In section IV, we generalize these considerations
to a numerical investigation of higher dimensional settings.
Finally, in section V, we summarize our findings and
present our conclusions.

\section{Model Equations and Mathematical Setup}

We consider a set of coupled DNLS equations
\begin{eqnarray}\label{dyn1}
    i\dot U_n+(g_{11}|U_n|^2+g_{12}|V_n|^2)U_n+C\Delta_D U_n &=& 0,
    \nonumber\\
    i\dot V_n+(g_{12}|U_n|^2+g_{22}|V_n|^2)V_n+C\Delta_D V_n &=& 0,
\end{eqnarray}
where $n$ is a $D$-Dimensional index and $\Delta_D$ is the discrete
Laplacian in $D$ dimensions. We look for stationary solutions
$\{u_n\}$, $\{v_n\}$ through the relation
\begin{equation}
    U_n(t)=\exp(i\Lambda_1t)u_n,\qquad V_n(t)=\exp(i\Lambda_2t)v_n.
\end{equation}

The dynamical equations (\ref{dyn1}) then transform into
\begin{eqnarray}\label{dyn2}
    -\Lambda_1 u_n+(g_{11}|u_n|^2+g_{12}|v_n|^2)u_n+C\Delta_D u_n &=& 0,
    \nonumber\\
    -\Lambda_2 v_n+(g_{12}|v_n|^2+g_{22}|v_n|^2)v_n+C\Delta_D v_n &=& 0.
\end{eqnarray}

The stability is determined in a frame rotating with frequency
$\Lambda_1$ for $U_n(t)$ and $\Lambda_2$ for $V_n(t)$, i.e., we
suppose that
\begin{equation}
    U_n(t)=\exp(i\Lambda_1t)[u_n+\xi^{(1)}_n(t)],\qquad
    V_n(t)=\exp(i\Lambda_2t)[v_n+\xi^{(2)}_n(t)].
\end{equation}

The small perturbations $\xi^{(k)}_n(t)$, with $k=1,2,$ can be
expressed as
\begin{equation}
    \xi^{(1)}_n(t)=a_n\exp(i\lambda t)+b_n\exp(-i\lambda^*t),\qquad
    \xi^{(2)}_n(t)=c_n\exp(i\lambda t)+d_n\exp(-i\lambda^*t),
\end{equation}
leading to the linear stability equations
\begin{equation}
\lambda J \overline{\xi_n} =
M_n \overline{\xi_n}
+C (\overline{\xi_{n+1}}+\overline{\xi_{n-1}}),
\label{evp}
\end{equation}
with
\begin{eqnarray}
\overline{\xi_n}&=&(a_n\quad b_n^* \quad c_n \quad d_n^*)^T,\quad J=\left(
\begin{array}{cccc}
1 & 0 & 0 & 0 \\
0 & -1 & 0 & 0\\
0 & 0 & 1 &0\\
0 & 0 & 0 & -1
\end{array}
\right),\\
M_n&=&\left(
\begin{array}{ccccc}
K_{1,n}& g_{11}u_n^2 & g_{12} u_nv_n^* & g_{12}u_nv_n\\
g_{11}(u_n^2)^* & K_{1,n} & g_{12} u_n^*v_n^* & g_{12}u_n^*v_n\\
g_{12} u_n^*v_n & g_{12}u_nv_n & K_{2,n} & g_{22}v_n^2\\
g_{12} u_n^*v_n^* & g_{12}u_nv_n^* & g_{22}(v_n^2)^* & K_{2,n}
\end{array}\right),\\
K_{1,n}&=&-\Lambda+2g_{11}|u_n|^2+g_{12}|v_n|^2-2C,\nonumber\\
K_{2,n}&=&-\Lambda+2g_{22}|v_n|^2+g_{12}|u_n|^2-2C.\nonumber
\end{eqnarray}

Soliton and vortex solutions are calculated using methods based on
the anti-continuous limit. Upon calculating these solutions
at $C=0$, we  continue them to finite coupling
by varying $C$ or other parameters (such as the interspecies
nonlinearity strength $g_{12}$).

We are interested in interlaced solitons (ISs) in 1D lattices and
interlaced vortices (IVs) in 2D and 3D lattices. The excited sites at $C=0$ are
equal to $\tilde{u}$ and $\tilde{v}$, except for a phase factor $\exp(i\phi)$,
while $u_nv_n=0$ at the corresponding excited site. These values are
\begin{equation}\label{ac0}
    \tilde{u}=0,\sqrt{\Lambda_1/g_{11}},\qquad \tilde{v}=0,\sqrt{\Lambda_2/g_{22}}.
\end{equation}

In what follows, we choose $\Lambda_1=\Lambda_2\equiv\Lambda$ and $g_{11}=g_{22}=1$. We also choose $g_{12}\leq1$ as, for $g_{12}>1$ interlaced solitons and vortices are unstable for every value of $C$.

\section{Analytical and Numerical Results for 1d Interlaced Solitons}

\subsection{Existence and stability}

We consider interlaced solitons which are labeled by
$|AB>\equiv|A>|B>$, where $A,B=0,1,2,\ldots$. This number indicates
the ``order'' of the excited state at the anti-continuous limit, whose phase
$\phi=0,\pi$ is chosen so that the isolated solitons (i.e. when
$g_{12}=0$) are stable for any small $C$.
For instance, the ground state $|0>$ means $u_n=\tilde{u}\delta_{n,0}$ and the
first excited
state $|1>$ will be taken to mean
$u_n=\tilde{u}(\delta_{n,1}-\delta_{n,-1})$ at the AC
limit. Thus, the state $|01>$ corresponds to $u_n=\tilde{u}\delta_{n,0}$,
$v_n=\tilde{v}(\delta_{n,1}-\delta_{n,-1})$ and $|12>$ to
$u_n=\tilde{u}(\delta_{n,1}-\delta_{n,-1})$,
$v_n=\tilde{v}(\delta_{n,2}+\delta_{n,-2})-\delta_{n,0}$.

\begin{figure}
\begin{center}
\begin{tabular}{ccc}
    (a) & (b) & (c)\\
    \includegraphics[width=4.5cm]{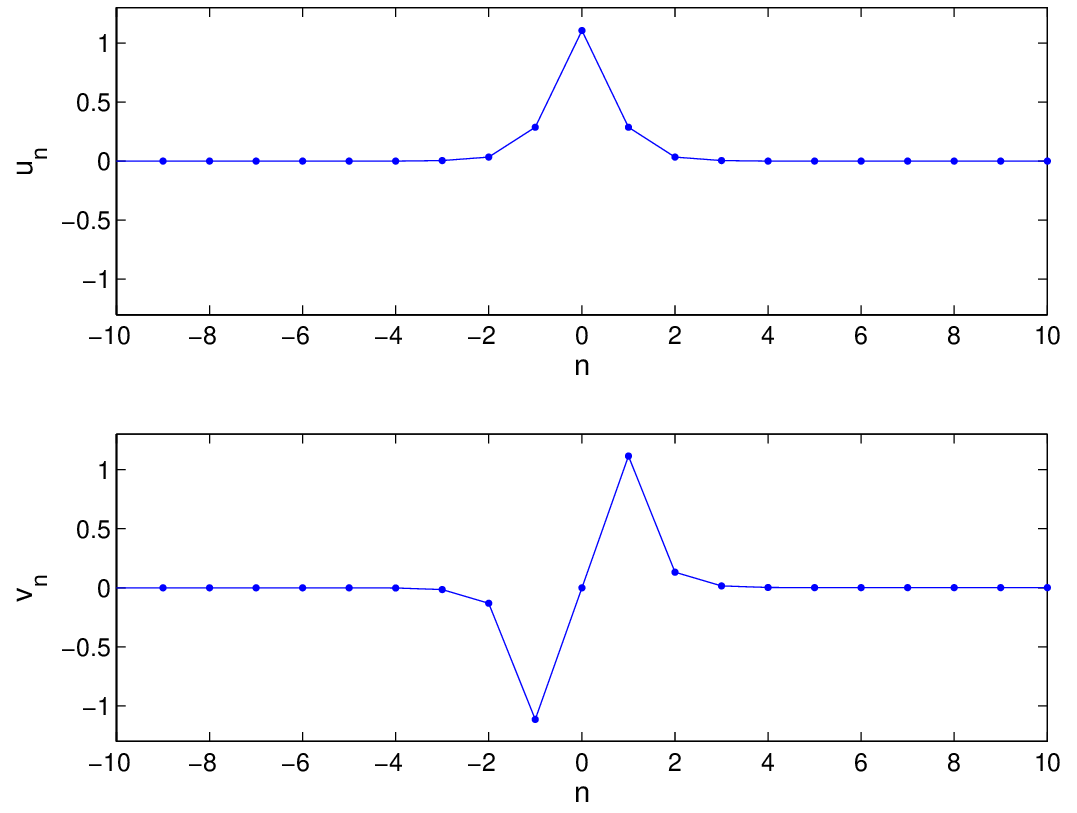} &
    \includegraphics[width=4.5cm]{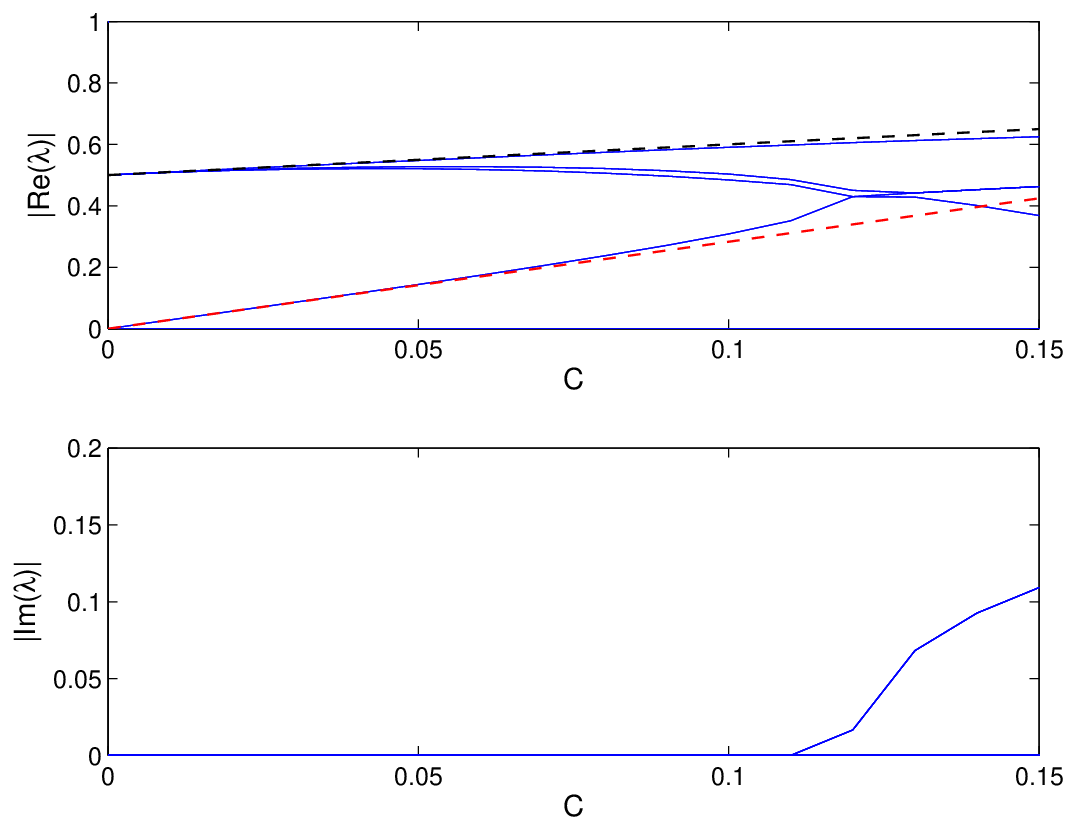} &
    \includegraphics[width=4cm]{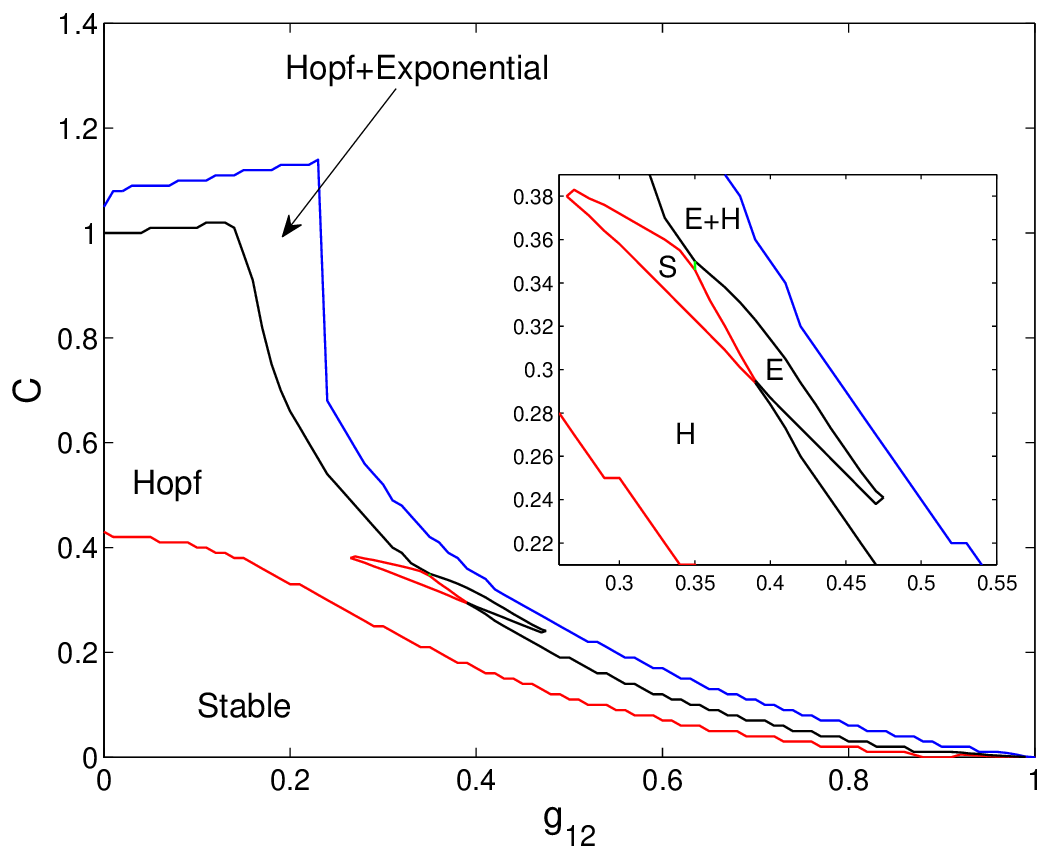} \\
\end{tabular}
\caption{(a) Profiles of $|01 \rangle$ interlaced solitons with $g_{12}=0.5$ and $C=0.15$. (b) Dependence on $C$ of the real and imaginary parts  of eigenfrequencies of small perturbations about $|01>$ with $g_{12}=0.5$. Dashed lines correspond to Lyapunov-Schmidt predictions of equations (\ref{eq:LS01}), and (\ref{eq:LS0}). (c) Two-parameter stability diagram in the plane of intersite ($C$) and inter-component ($g_{12}$) coupling, indicating regions of occurance of Hopf bifurcations (H), exponential instability (E), and stability domain (S).} \label{fig1}
\end{center}
\end{figure}

We first analyze the $|01>$ state, which is stable for
$C<C_0$. At $C=C_0$ the ISs become unstable through Hopf bifurcations
(the value of $C_0$ differs as a function of the rest of the system
parameters such as $g_{12}$, however the above scenario is robust).
Cascades of this type of bifurcations arise as $C$ increases and, when,
$C\ge C_1$, the ISs become also exponentially unstable. There is a special
region for $g_{12}\in[0.27,0.37]$ where the system experiences an inverse Hopf
bifurcation recovering the stability in a window. The system becomes
unstable again through Hopf bifurcations for $g_{12}\in[0.27,0.34]$ and
exponential instabilities for $g_{12}\in[0.35,0.37]$. Besides, for
$g_{12}\in[0.38,0.47]$ there exist windows with only exponential
instabilities. Fig. \ref{fig1} illustrates all of the above features, by showcasing
a typical example of the $|01>$ state, a typical continuation of
its principal linearization eigenfrequencies $\lambda$, and a full
two-parameter diagram of the stability of this state in the
two-parameter plane $(C,g_{12})$.

\begin{figure}
\begin{center}
\begin{tabular}{ccc}
    (a) & (b) & (c)\\
    \includegraphics[width=4.5cm]{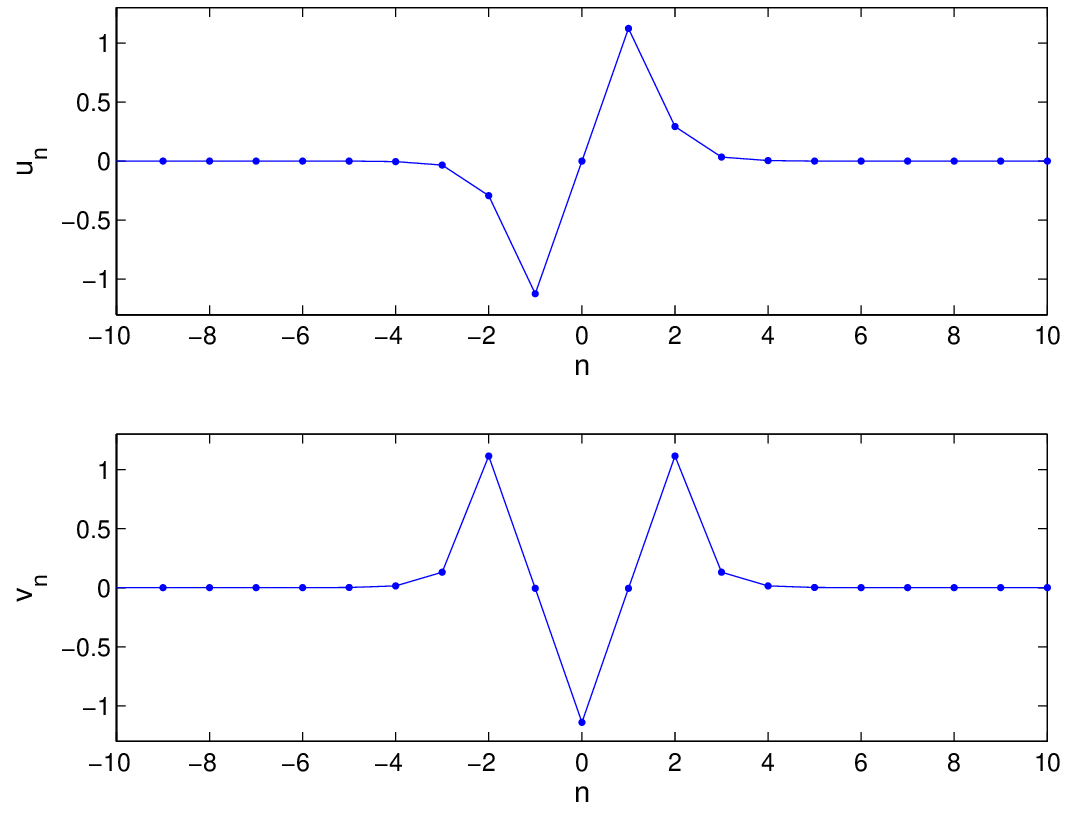} &
    \includegraphics[width=4.5cm]{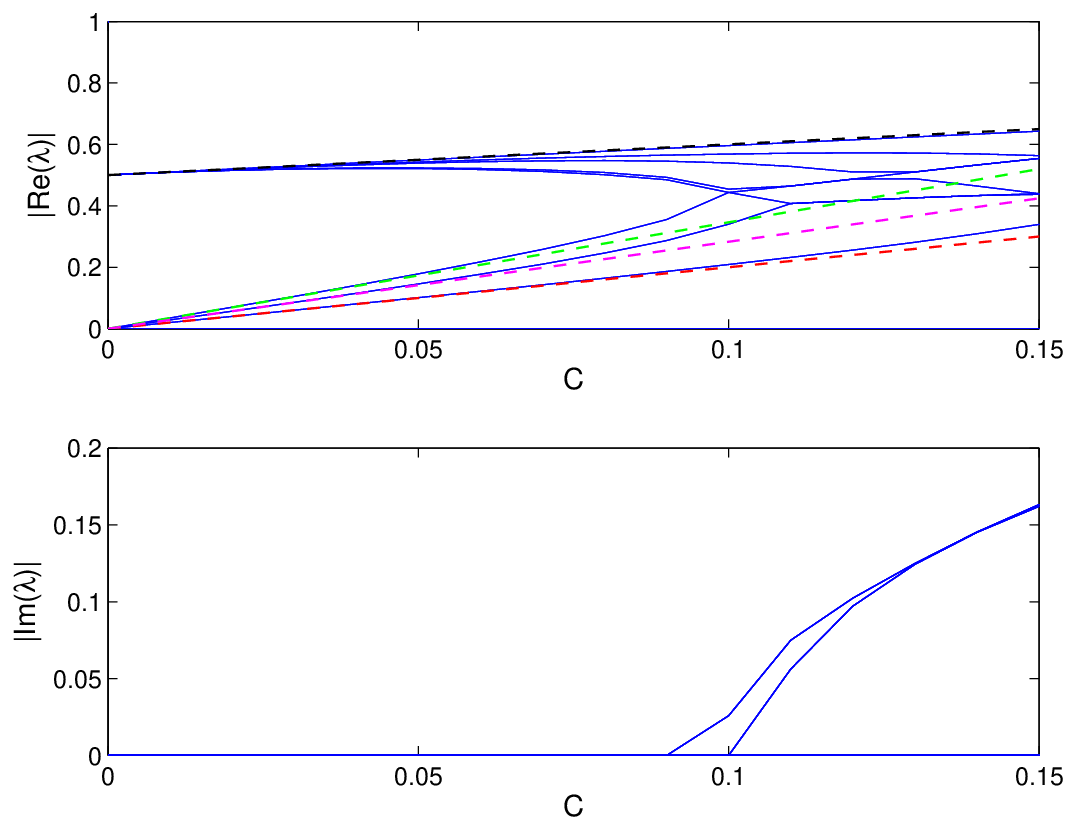} &
    \includegraphics[width=4cm]{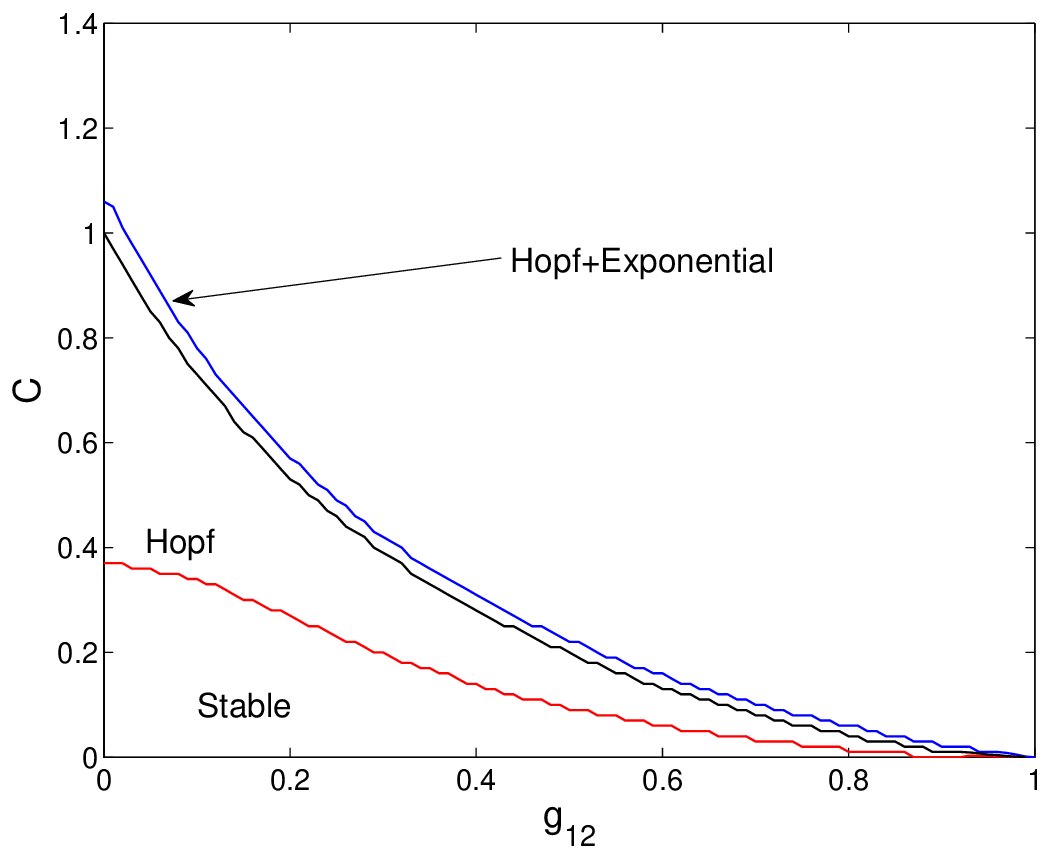} \\
\end{tabular}
\caption{(a) Profiles and (b) dependence on $C$ of the real and imaginary parts of eigenfrequencies of small perturbations of $|12 \rangle$ showing the same features and for the same parameters as in Fig \ref{fig1}. Dashed lines correspond to Lyapunov-Schmidt predictions of equations (\ref{eq:LS12a}), and (\ref{eq:LS0}). (c) Two-parameter stability diagram in the plane of intersite ($C$) and inter-component ($g_{12}$) coupling.}\label{fig2}
\end{center}
\end{figure}

For $|12>$ states, the scenario is essentially similar to the $|01>$ case,
although, in essence, it is considerably simpler due to the absence
of any inverse Hopf bifurcations and restabilization windows.
Fig. \ref{fig2} shows the corresponding features for
$|12>$, as Fig. \ref{fig1} for the $|01>$ case.

\subsection{Dynamics of unstable solitons}

\begin{figure}
\begin{center}
\begin{tabular}{cc}
    \includegraphics[width=4.5cm]{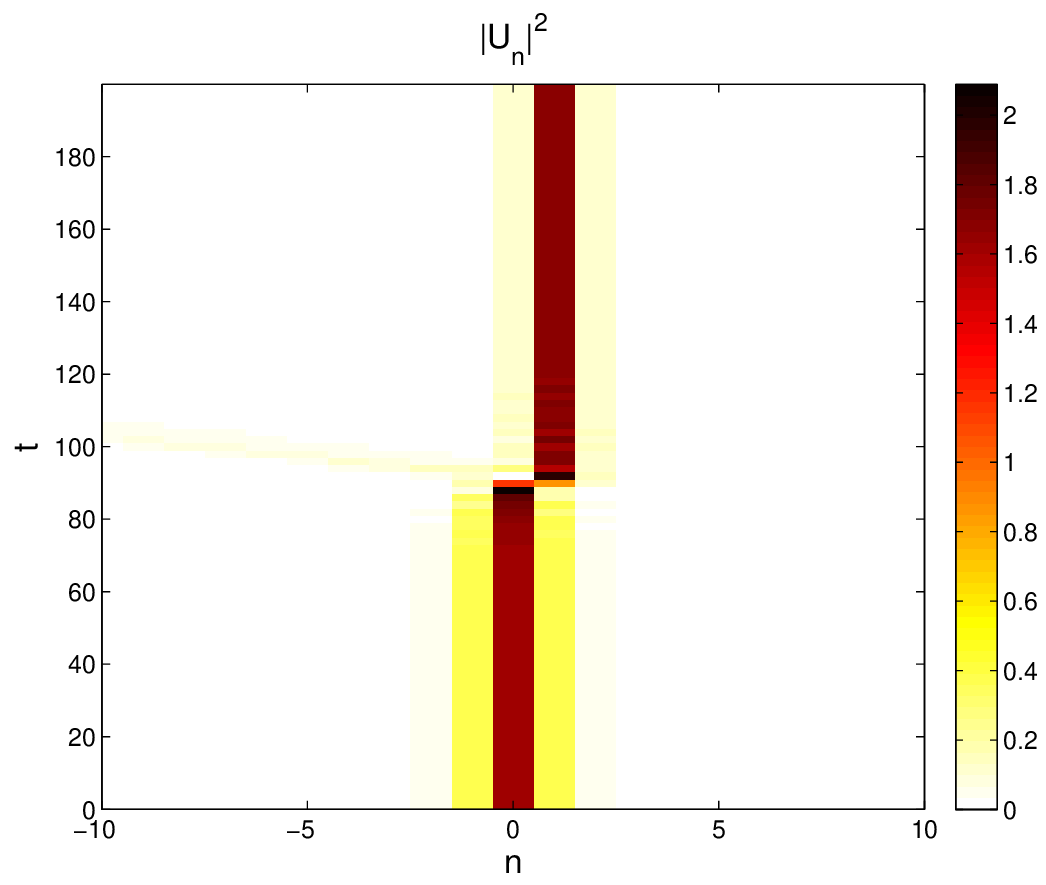} &
    \includegraphics[width=4.5cm]{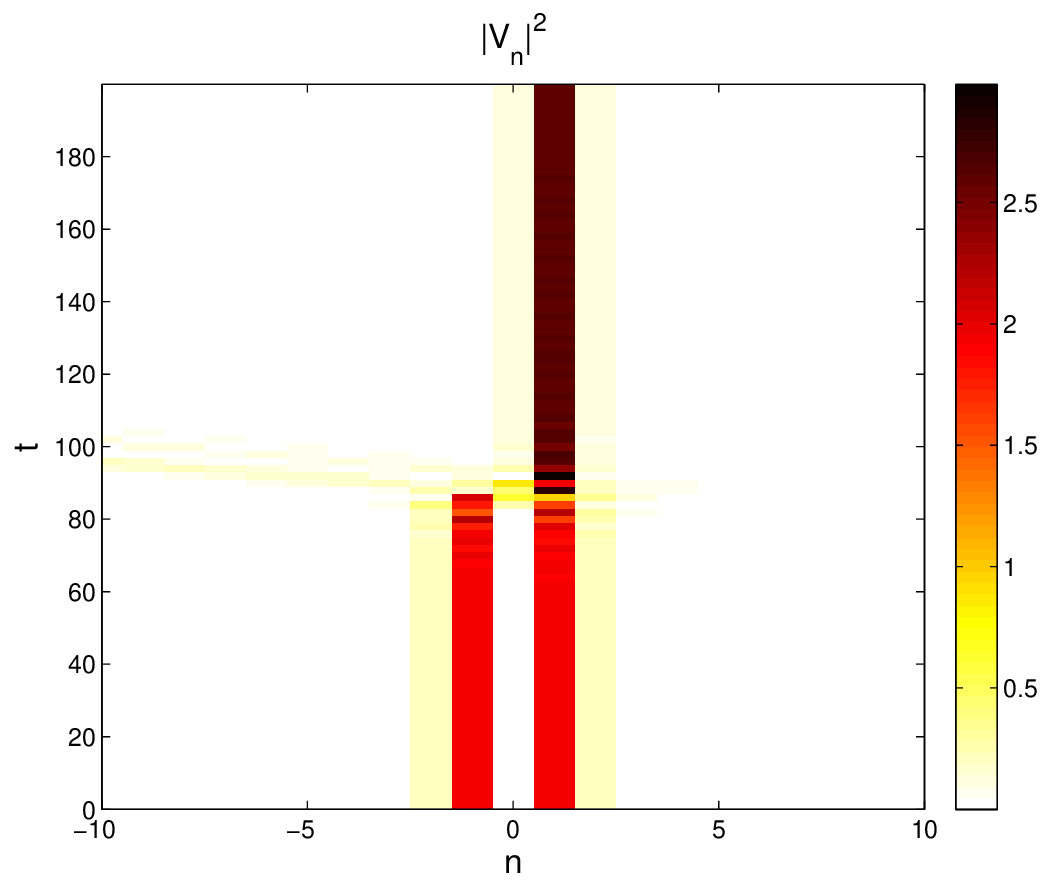} \\
\end{tabular}
\caption{Time evolution of the density of the two components for
a slightly perturbed unstable $|01>$ IS with $g_{12}=0.2$ and $C=0.6$.}
\label{fig3}
\end{center}
\end{figure}

\begin{figure}
\begin{center}
\begin{tabular}{cc}
    \includegraphics[width=4.5cm]{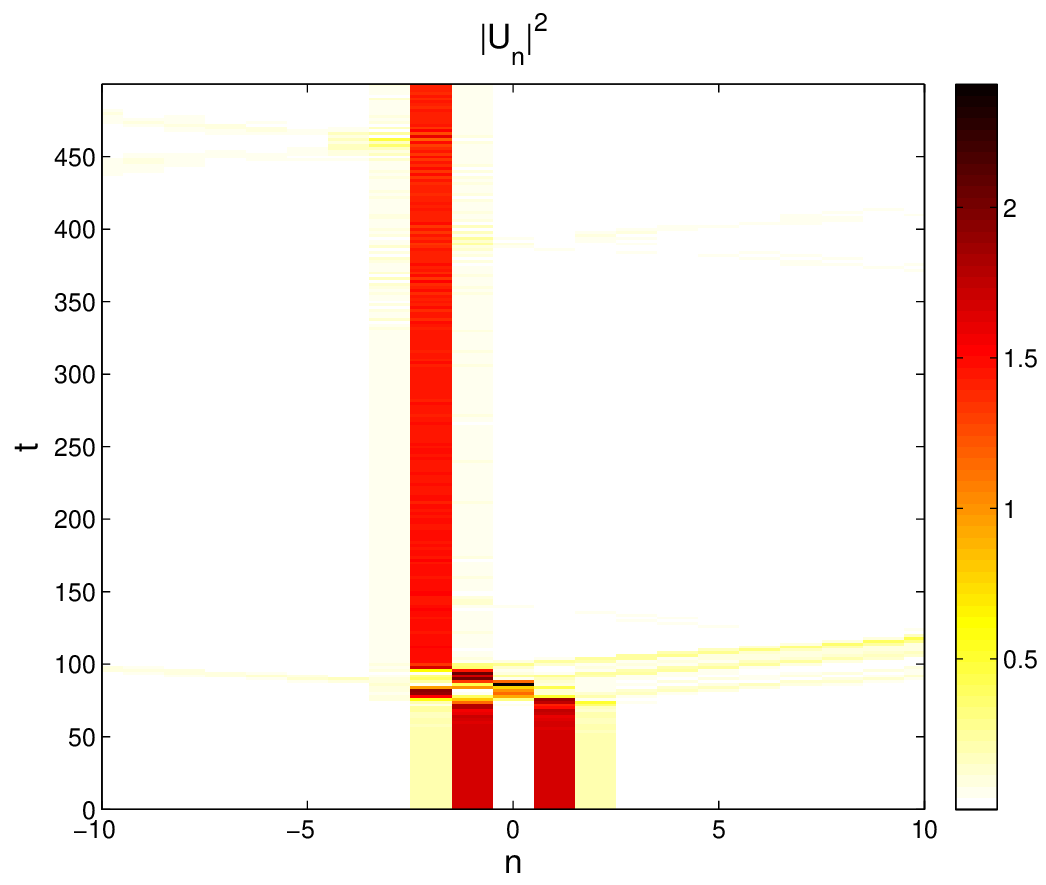} &
    \includegraphics[width=4.5cm]{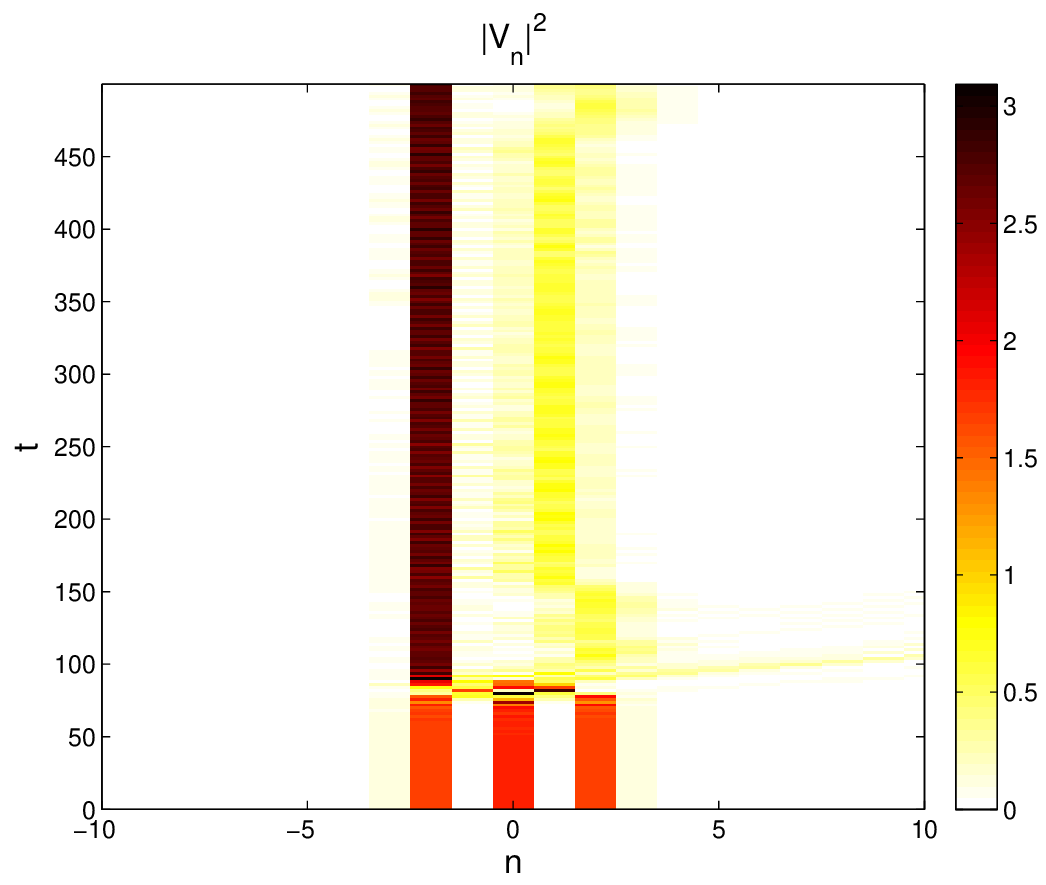} \\
\end{tabular}
\caption{Time evolution of the density of the two components for
a slightly perturbed unstable $|12>$ IS with $g_{12}=0.2$ and $C=0.4$.}
\label{fig4}
\end{center}
\end{figure}

First, we analyze the dynamics of $|01>$ ISs. Fig. \ref{fig3} shows the evolution of a typically unstable (i.e. oscillatory unstable) $|01>$ IS with $g=0.2$ and $C=0.6$. The oscillatory evolution of the instability
eventually transforms the mode into a $|00>$ state, which is a stable
state of the system.
The final excited site is typically the same for the $\{U_n\}$ and $\{V_n\}$
coordinates, although in some cases (even for the same parameters set), the
asymptotic excited site does not need to be same. However, the amplitude of
for the $n$th site is not identical, i.e. $|u_n|\neq |v_n|$. In a similar
vein,
Fig. \ref{fig4} shows the evolution of an oscillatory unstable $|12>$ IS with
$g=0.2$ and $C=0.4$, and, analogously to the $|01>$ case, the IS
evolves to a $|00>$ state (although the finally populated site is
not the central one of the original configuration).

\subsection{Perturbation analysis}

In this subsection, we attempt to understand in some more details
the above observed results of the numerical computations in
connection to the stability properties of the interlaced soliton
solutions. More specifically, we evaluate explicit
expressions of the interlaced solitons' eigenvalues for the
configurations discussed above.
The method is based on the expansion in the coupling constant $C$,
in the vicinity of the anti-continuum limit.

In the limit $C=0$, as illustrated above, there are two types
of solutions, i.e.\ $u_n=v_n=0$, and the non-zero solutions given by
Eqs.\ (\ref{ac0}). In this limit, one can also easily
notice that the eigenvalue problem (\ref{evp}) will give
\begin{equation}
\lambda=\pm\Lambda,\,\pm\Lambda(1-g_{12}/g_{11}),\,\pm\Lambda(1-g_{12}/g_{22})
\label{ev_C0_1}
\end{equation}
for the zero solutions and
\begin{equation}
\lambda=\pm0
\label{ev_C0_2}
\end{equation}
for the non-zero solutions (\ref{ac0}).

It can be directly inferred from the analysis of the underlying linear problem
that the stable eigenvalues $\lambda=\pm\Lambda$ will expand creating a band of
continuous spectrum when $C$ is increased. Therefore, this eigenvalue will not be
discussed further. The instability for a soliton solution will then be determined
by the bifurcation of the remaining eigenvalues.

Let us now first consider the profile of $|01>$ ISs. It is clear that
for finite $C$ the solutions will be deformed from their
AC-limit profile. The leading-order solution up to ${\rm O}(C)$ is then found to be
\begin{equation}
\begin{array}{lll}
\displaystyle &&u_0=\sqrt{\frac{\Lambda}{g_{11}}}+\frac{C}{\sqrt{\Lambda g_{11}}},\quad u_1=u_{-1}=\frac{C}{\sqrt{\Lambda g_{11}(1-g_{12}/g_{22})}},\\
\displaystyle &&v_0=0,\quad v_1=-v_{-1}=\sqrt{\frac{\Lambda}{g_{22}}}+\frac{C}{\sqrt{\Lambda g_{22}}}.
\end{array}
\end{equation}

The next step is to consider
the stability problem when the coupling is turned on. To the leading order, the eigenvalue problem of this particular configuration is then given by
\begin{equation}
\mathcal{M}\, \Xi=\lambda\, \sigma\, \Xi,
\label{evp2}
\end{equation}
where
\begin{equation}
\sigma=\textrm{diag}(J)
, \quad
\Xi=\left(
\begin{array}{cccc}
\overline{\xi_{-2}}\\
\overline{\xi_{-1}}\\
\overline{\xi_{0}}\\
\overline{\xi_{1}}\\
\overline{\xi_{2}}
\end{array}\right),
\quad
\mathcal{M}=\left(
\begin{array}{cccccc}
M_{-2} & C Id_{4\times4} & 0 & 0 & 0\\
C Id_{4\times4} & M_{-1} & C Id_{4\times4} & 0 & 0 \\
0 & C Id_{4\times4} & M_{0} & C Id_{4\times4} & 0\\
0&0 & C Id_{4\times4} & M_{1} & C Id_{4\times4}\\
0&0&0 & C Id_{4\times4} & M_{2}
\end{array}
\right)
,
\end{equation}
and $Id_{4\times4}$ is the identity matrix of size ${4\times4}$.

Since we have expanded
$u_n$ and $v_n$ in a power series of $C$, then it is natural that
we also expand all the involved quantities in $C$, i.e.\  $\mathcal{M}=\mathcal{M}_{0}+C\mathcal{M}_{1}+C^2\mathcal{M}_{2}+\mathcal{O}(C^3)$, $\Xi=\Xi_0+C\Xi_1+C^2\Xi_2+{\mathcal O}(C^3)$ and $\lambda={{\lambda_0}+C\lambda_1+C^2\lambda_2+{\mathcal O}(C^3)}$. It can be checked that $\mathcal{M}_0$ is a singular self-adjoint matrix.

Substituting the expansions to the eigenvalue problems (\ref{evp2}) will give us to the leading order
\begin{equation}
\mathcal{M}_{0}\,\Xi_0=\lambda_0\,\sigma\,\Xi_0,
\end{equation}
from which one will obtain that $\lambda_0$ is given by Eqs.\ (\ref{ev_C0_1})
and (\ref{ev_C0_2}). In the following, let us first consider the case of $\lambda_0=0$ which are of three pairs, with the corresponding eigenvalues of $\mathcal{M}_{0}\,\Xi_0=0$ denoted by $e_j,\,j=1,2,3$. Therefore, one can write
\[\Xi_0=\sum_{j=1}^3 c_j\,e_j.\]

The next order equation of (\ref{evp2}) gives us
\begin{equation}
\mathcal{M}_{0}\,\Xi_1=\lambda_1\,\sigma\,\Xi_0-\mathcal{M}_1\,\Xi_0.
\label{evp3}
\end{equation}
Using the Fredholm alternative theorem, the above equation will have a solution if
the right hand side is orthogonal to the null space of $\mathcal{M}_0$, which it is. Hence, the value of the correction $\lambda_1$ cannot be obtained yet and a solution $\Xi_1$ of (\ref{evp3}) can therefore be calculated for \emph{any} $\lambda_1$.

The equation of order $\mathcal{O}(C^3)$ from (\ref{evp}) can be easily deduced
to be
\begin{equation}
\mathcal{M}_{0}\,\Xi_2=\lambda_2\,\sigma\,\Xi_0+\lambda_1\,\sigma\,\Xi_1-\mathcal{M}_1\,\Xi_1-\mathcal{M}_2\,\Xi_0.
\label{evp4}
\end{equation}
Projecting the equation above to $e_j$, $j=1,2,3$, i.e.\ basis of the null space of $\mathcal{M}_{0}$, will give us the following eigenvalue matrix
\begin{equation}
\left(
\begin{array}{cccccc}
\frac{-2g_{11}}{(g_{11}-g_{12})\Lambda} & 0 & \frac{-2g_{11}}{(g_{11}-g_{12})\Lambda}\\
0 & 0 & 0 \\
\frac{-2g_{11}}{(g_{11}-g_{12})\Lambda} & 0 & \frac{-2g_{11}}{(g_{11}-g_{12})\Lambda}
\end{array}
\right)\,
\left(\begin{array}{c} c_1\\c_2\\c_3 \end{array}\right)
=-\frac{\lambda_1^2}{\Lambda}\,\left(\begin{array}{c} c_1\\c_2\\c_3 \end{array}\right),
\label{conf1}
\end{equation}
which can be immediately solve to yield
\begin{equation}
\lambda_1=\pm0,\,\pm0,\,\pm2\sqrt{\frac{g_{11}}{g_{11}-g_{12}}}.
\end{equation}
This illustrates that there is a pair of eigenvalues bifurcating
from zero as given by
\begin{equation}\label{eq:LS01}
\lambda=\pm2C\sqrt{\frac{g_{11}}{g_{11}-g_{12}}}+\mathcal{O}(C^2).
\end{equation}

The same procedure can be applied to bifurcations of the non-zero eigenvalues.
In this case, the calculation is even simpler as applying the Fredholm alternative
to the $\mathcal{O}(C)$ equation of (\ref{evp2}) already gives us a solvability
condition from which we obtain that bifurcating eigenvalues are
\begin{equation}\label{eq:LS0}
\lambda=\pm(1-g_{12}/g_{22})(\Lambda+2C),\,\pm(1-g_{12}/g_{11})(\Lambda+2C),
\end{equation}

The above procedure can also be similarly and immediately applied to the
configuration $|12>$ ISs. 
The only difference is that for that solution one will obtain a
stability matrix $\mathcal{M}$ of size $28\times28.$

For $|12>$ISs, we can obtain the solution in a power series of $C$ as
\begin{equation}
\begin{array}{lll}
&&u_0=0,\quad u_1=-u_{-1}=\sqrt{\frac{\Lambda}{g_{11}}}+\frac{C}{\sqrt{\Lambda g_{11}}},\quad u_2=-u_{-2}=\frac{C}{\sqrt{\Lambda g_{11}}(1-g_{12}/g_{22})},\\
&&v_0=-v_2=-v_{-2}=-\sqrt{\frac{\Lambda}{g_{22}}}-\frac{C}{\sqrt{\Lambda g_{22}}},\quad v_1=v_{-1}=0.
\end{array}
\end{equation}

Continuing to finding the eigenvalues, we will also immediately obtain that in place of (\ref{conf1}), one will obtain the following eigenvalue problem
\begin{equation}
\left(
\begin{array}{cccccc}
\frac{-2g_{11}}{(g_{11}-g_{12})\Lambda} & 0 & \frac{-2g_{11}}{(g_{11}-g_{12})\Lambda} & 0 & 0\\
0 & \frac{-2g_{22}}{(g_{22}-g_{12})\Lambda} & 0 & \frac{-2g_{22}}{(g_{22}-g_{12})\Lambda} & 0 \\
\frac{-2g_{11}}{(g_{11}-g_{12})\Lambda} & 0 & \frac{-4g_{11}}{(g_{11}-g_{12})\Lambda} & 0 & \frac{-2g_{11}}{(g_{11}-g_{12})\Lambda}\\
0 & \frac{-2g_{22}}{(g_{22}-g_{12})\Lambda} & 0 & \frac{-2g_{22}}{(g_{22}-g_{12})\Lambda} & 0 \\
0 & 0 & \frac{-2g_{11}}{(g_{11}-g_{12})\Lambda} & 0 & \frac{-2g_{11}}{(g_{11}-g_{12})\Lambda}
\end{array}
\right)\,
\left(\begin{array}{c} c_1\\c_2\\c_3\\c_4\\c_5\end{array}\right)=-\frac{\lambda_1^2}{\Lambda}\,\left(\begin{array}{c} c_1\\c_2\\c_3\\c_4\\c_5\end{array}\right),
\label{conf2}
\end{equation}
from which we can obtain eigenvalues bifurcating from zero as
\begin{equation}\label{eq:LS12a}
\lambda=\pm\sqrt{\frac{2}{1-g_{12}/g_{11}}}C,\,\pm\sqrt{\frac{6}{1-g_{12}/g_{11}}}C,\,\pm\sqrt{\frac{4}{1-g_{12}/g_{22}}}C.
\end{equation}
Bifurcations from the non-zero eigenvalues for this case can also be shown to
yield Eq.\ (\ref{eq:LS0}).

The above analytical expressions give us a detailed handle on the dependence
of the relevant eigenalues on the system parameters.
Comparisons of the analytical results obtained here with the numerical ones are
presented in Figs.\ \ref{fig1}-\ref{fig2} where one can see that the analytical
expressions are in relatively good agreement with the numerical results.
It should be noted that although such analytical considerations are
procedurally straightforward to generalize in higher dimensions, the
relevant calculations are extremely tedious and will thus not be
pursued here. Instead, we now turn to numerical computations to
showcase the existence and potential stability of interlaced solitons
and vortices in higher-dimensional settings.

\section{Numerical Results for Interlaced Structures in Higher Dimensions}

For the case of 2D lattices, we consider two different interlaced structures.
On the one hand, we examine interlaced vortices (IVs) whose configurations in the AC limit are given by $u_{0,1}=\tilde{u}$, $u_{0,-1}=-\tilde{u}$, $u_{1,0}=i\tilde{u}$, $u_{-1,0}=-i\tilde{u}$; $v_{1,1}=-i\tilde{v}$, $v_{1,-1}=\tilde{v}$, $v_{-1,-1}=i\tilde{v}$, $v_{-1,1}=-\tilde{v}$. On the other hand, we also study a discrete
soliton interlaced with a vortex (IVSs) whose configurations in the AC limit are
given by $u_{0,1}=\tilde{u}$, $u_{0,-1}=-\tilde{u}$, $u_{1,0}=i\tilde{u}$, $u_{-1,0}=-i\tilde{u}$, $v_{0,0}=1$.

\begin{figure}
\begin{center}
\begin{tabular}{ccc}
    (a) & (b) & (c) \\
    \includegraphics[width=4.5cm]{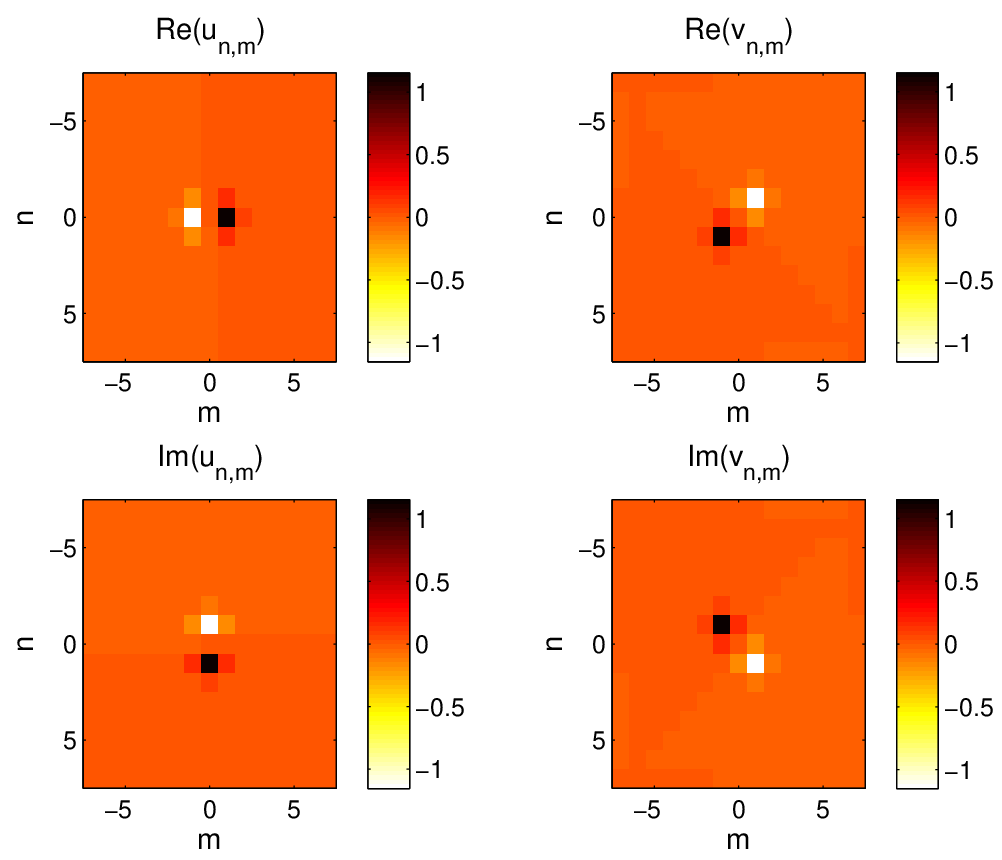} &
    \includegraphics[width=4.5cm]{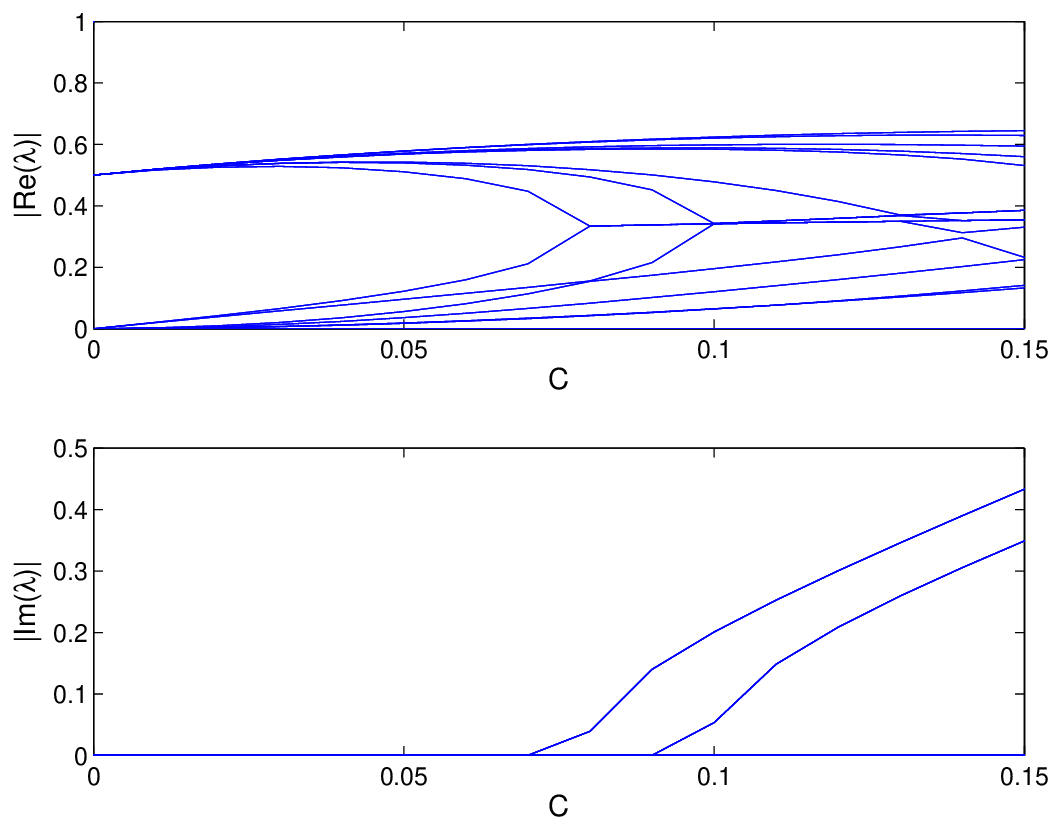} &
    \includegraphics[width=4cm]{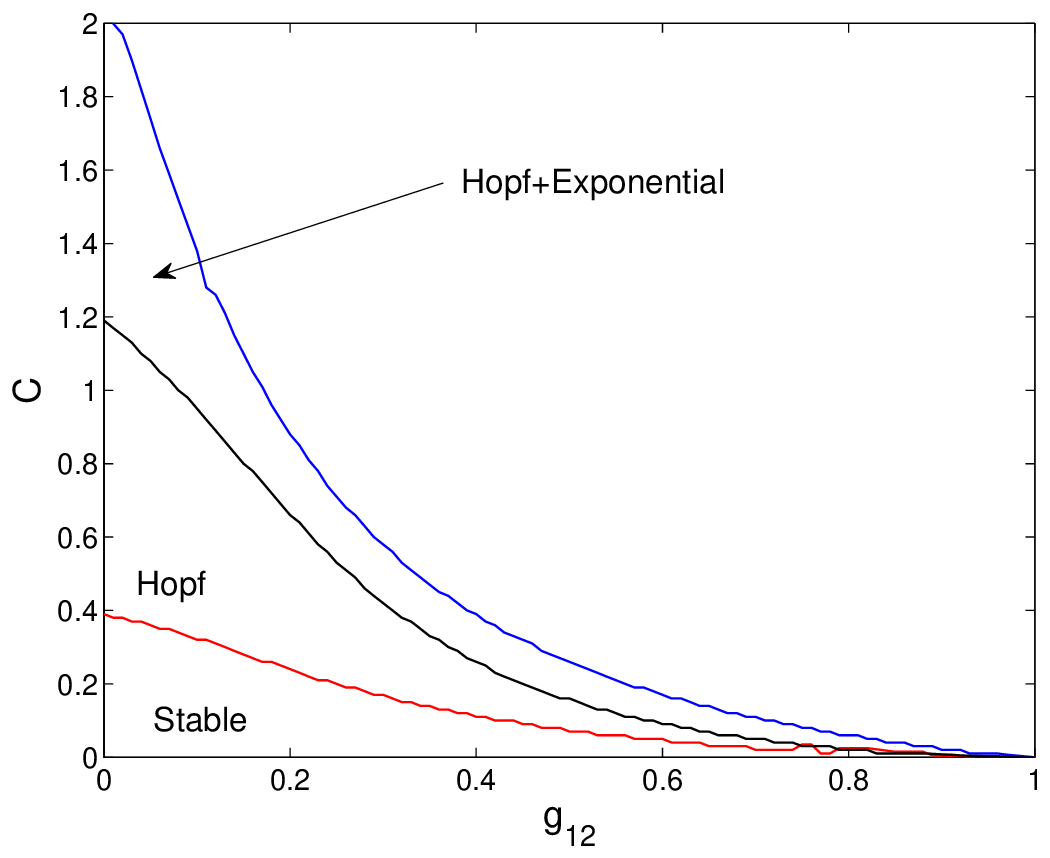} \\
\end{tabular}
\caption{(a) Plot of real (top) and imaginary (bottom) parts of interlaced
vortices with $g_{12}=0.5$ and $C=0.1$. (b) Dependence on $C$ of the real and
imaginary parts  of eigenfrequencies of small perturbations about such solutions
with $g_{12}=0.5$ (b) Two-parameter stability diagram in the plane of
intersite ($C$) and inter-component ($g_{12}$) coupling.}
\label{fig5}
\end{center}
\end{figure}

\begin{figure}
\begin{center}
\begin{tabular}{ccc}
    (a) & (b) & (c) \\
    \includegraphics[width=4.5cm]{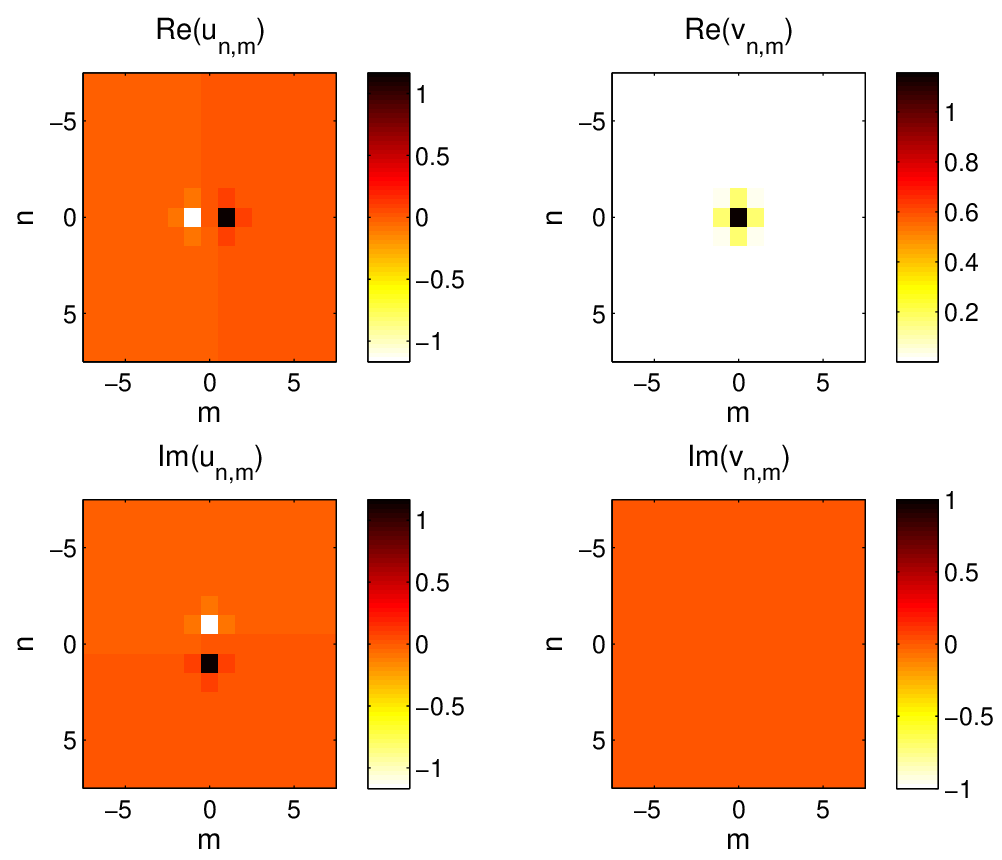} &
    \includegraphics[width=4.5cm]{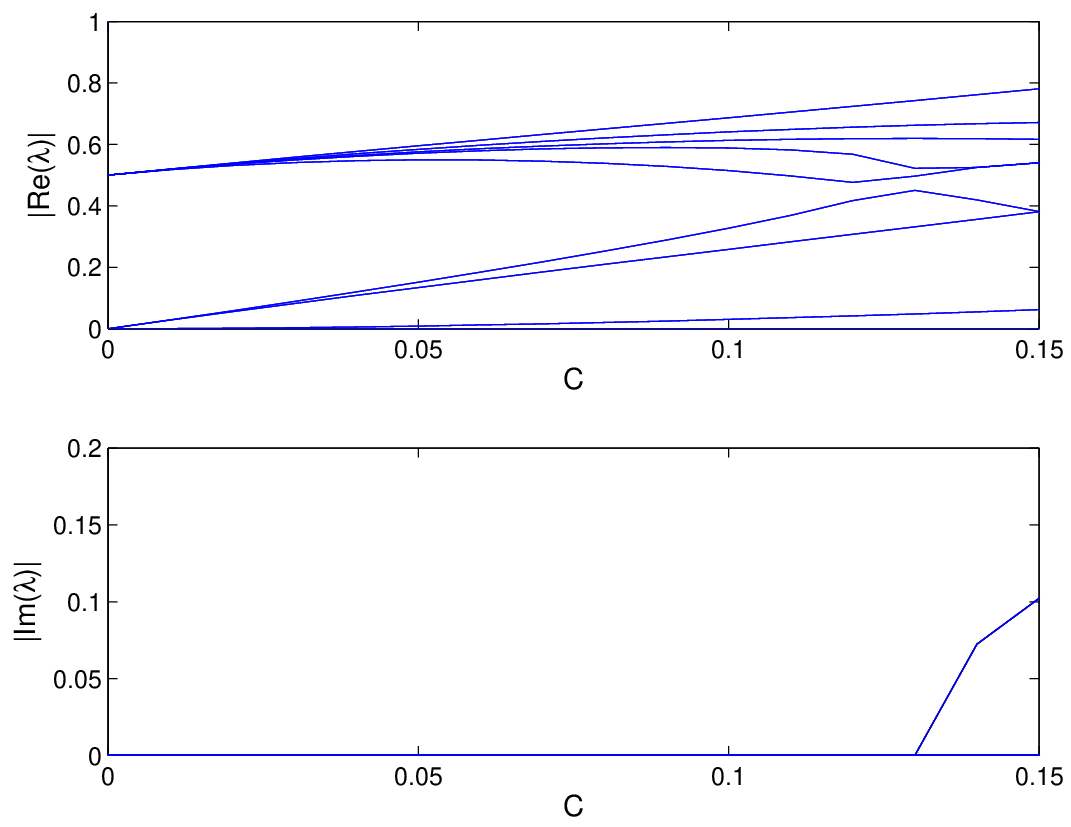} &
    \includegraphics[width=4cm]{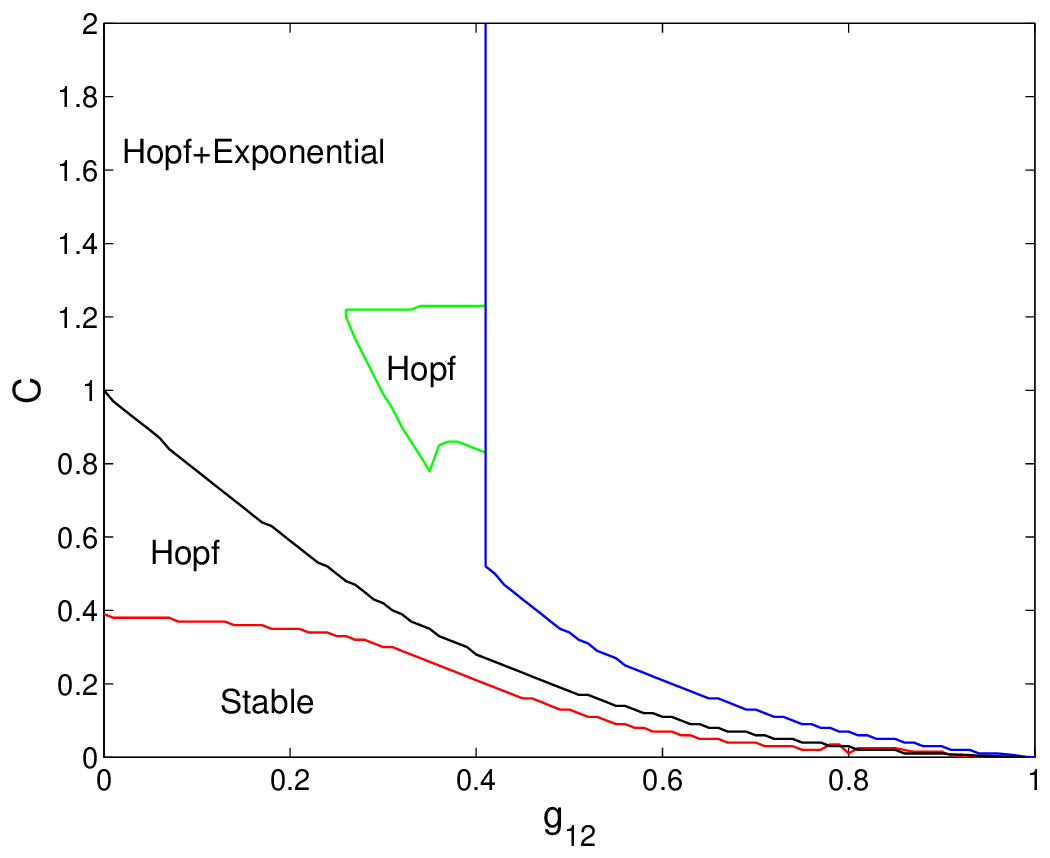} \\
\end{tabular}
\caption{(a) Plot of real (top) and imaginary (bottom) parts and (b) dependence
on $C$ of the real and imaginary parts  of eigenfrequencies of small perturbations
about interlaced soliton-vortex solutions, showing the same features and for the
same parameters as in Fig. \ref{fig5}. (c) The corresponding
two-parameter stability diagram.} \label{fig6}
\end{center}
\end{figure}

IVs experience a set of bifurcation scenaria which are qualitatively
similar to those of the $|12>$ ISs. IVSs experience the same scenario
as well, with the basic difference that they appear to exist for all $C$'s
(within the range examined i.e., up to $C=2$) for $g\leq0.4$. Also,
notably, the IVSs experience solely Hopf bifurcations in a fairly
small region inside the exponential+Hopf region. Figs. \ref{fig5} and \ref{fig6}
summarize the corresponding findings in a way similar as for the
1d configurations, presenting not only typical profiles of the modes,
but also typical mono-parametric continuations, as well as their full
two-parameter stability diagram in the space of inter-site and inter-species
coupling.

\begin{figure}
\begin{center}
\begin{tabular}{cc}
    (a) & (b) \\
    \includegraphics[width=6.6cm,height=5cm,angle=0]{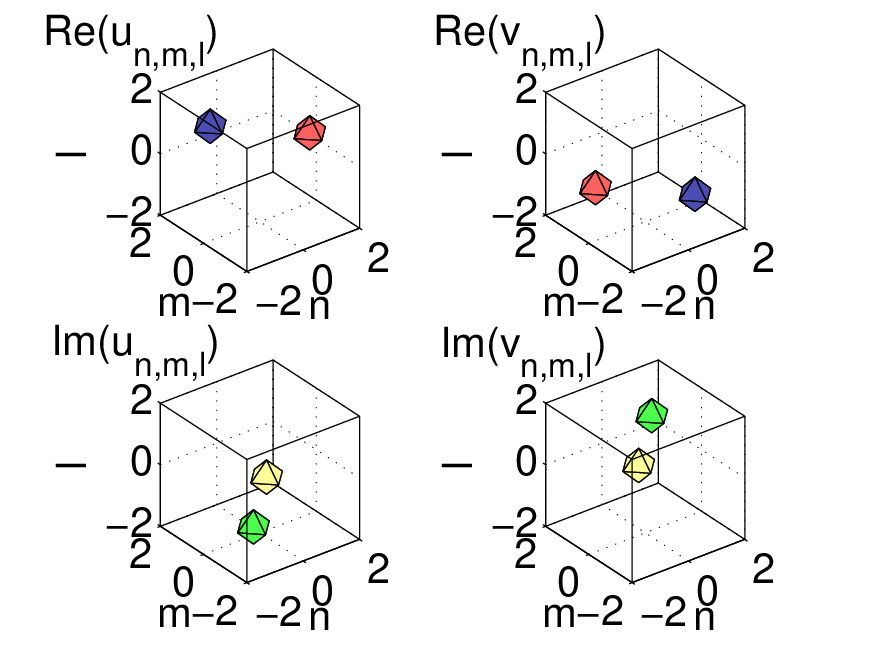} &
    \includegraphics[width=6.6cm,height=5.3cm,angle=0]{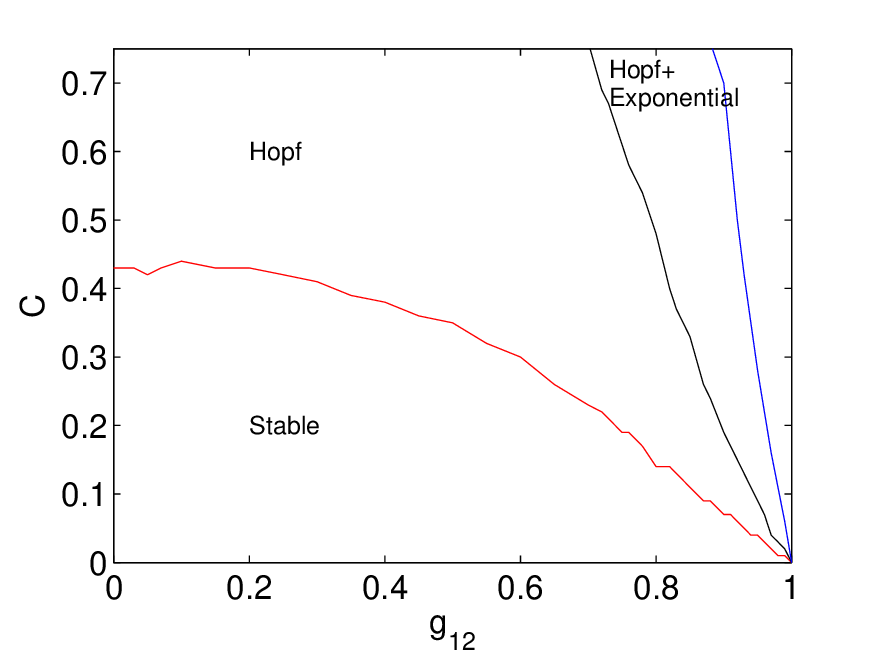} \\
    \includegraphics[width=6cm,height=2cm,angle=0]{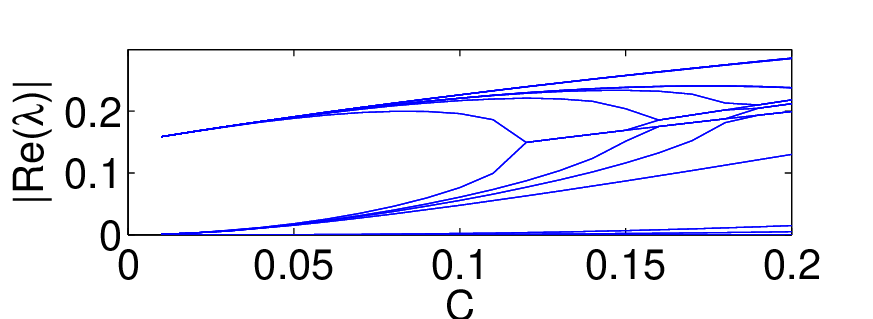} &
    \includegraphics[width=6.5cm,height=2cm,angle=0]{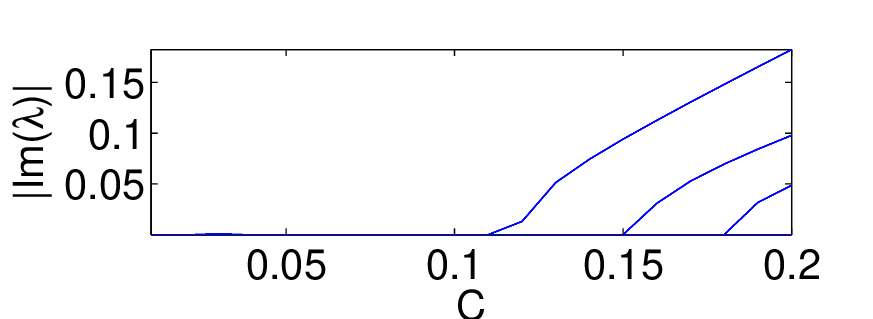}
\end{tabular}
\caption{(a) The top set of four panels show an interlaced cube ($g = 0.85$)
in a grid of size $11 \times 11 \times 11$ that has been continued to coupling $C = 0.20$. The level
contours shown correspond to $\mathrm{Re}(u_{n,m,l})=\mathrm{Re}(v_{n,m,l})=\pm 0.5 \max\left\{u_{n,m,l}\right\}$,
in blue and red (dark gray and gray, in the black-and-white version) respectively,
while the imaginary ones, $\mathrm{Im}(u_{n,m,l})=\mathrm{Im}(v_{n,m,l})=\pm 0.5\max\left\{u_{n,m,l}\right\}$,
are shown by green and yellow (light and very light gray, in the black-and-white version)
respectively. The bottom panel shows the real eigenfrequencies of small perturbation.
(b) The top panel shows the stability diagram, while the bottom panel shows the imaginary
eigenfrequencies of small perturbation.} \label{fig7}
\end{center}
\end{figure}

In the case of 3D lattices, we consider two interlaced vortices conjoined in the shape of a cube. In the AC limit,
this cube is given by $u_{-1,1,1} = \tilde{u}$, $u_{1,-1,1} = -\tilde{u}$, $u_{-1,-1-1} = i\tilde{u}$,
$u_{1,1,-1} = -i\tilde{u}$; $v_{1,-1,-1} = \tilde{v}$, $v_{-1,1,-1} = -\tilde{v}$, $v_{1,1,1} = i\tilde{v}$,
$v_{-1,-1,1} = -i\tilde{v}$. The structure is stable near the AC limit, with the size of the
window of stability diminishing as $g_{12}$ approaches 1, and with instability setting in via
Hopf bifurcations. In the 2-parameter continuation figure shown in Fig.~\ref{fig7}, the
coupling is only continued to $C = 0.75$, but it is observed that for values between
$g_{12} \approx 0.703$ and $g_{12} = 1$, the instability further degenerates into Hopf and exponential
instabilities. It should also be noted that within the region of instability,
there some exist isolated points or very narrow regions
where inverse Hopf bifurcations may be observed, which have been omitted
from the graph for clarity.
Let us note in passing here that the interlaced vortices in the ``vortex cube''
shown in Fig.~\ref{fig7} are perhaps not the prototypical interlaced structure
that one would expect in 3D; instead one might expect a structure where each
vortex is confined in a diagonal plane within the cube (with the two such planes
intersecting transversally). We were, however, unable to trace such a structure
even in the vicinity of the anti-continuum limit.

\subsection{Dynamics of unstable structures}

\begin{figure}
\begin{center}
\begin{tabular}{cc}
    (a) & (b) \\
    \includegraphics[width=4.5cm]{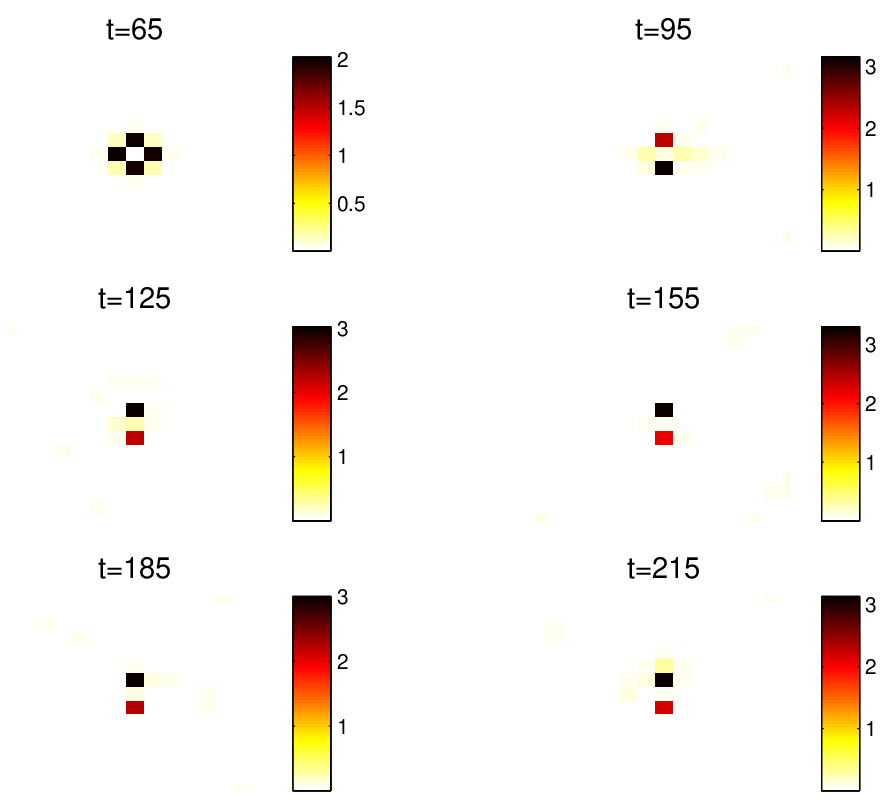} &
    \includegraphics[width=4.5cm]{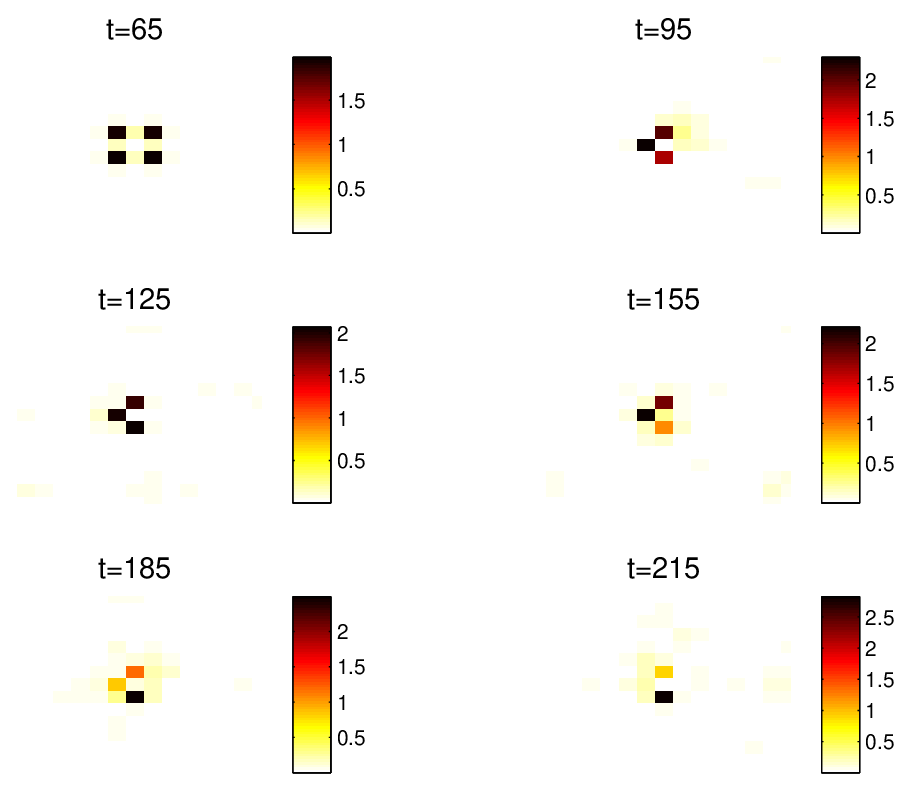} \\
\end{tabular}
\caption{Snapshots showing (a) $|U_n(t)|^2$ and (b) $|V_n(t)|^2$ for unstable IVs with $g_{12}=0.2$ and $C=0.3$.}
\label{fig8}
\end{center}
\end{figure}

\begin{figure}
\begin{center}
\begin{tabular}{cc}
    (a) & (b) \\
    \includegraphics[width=4.5cm]{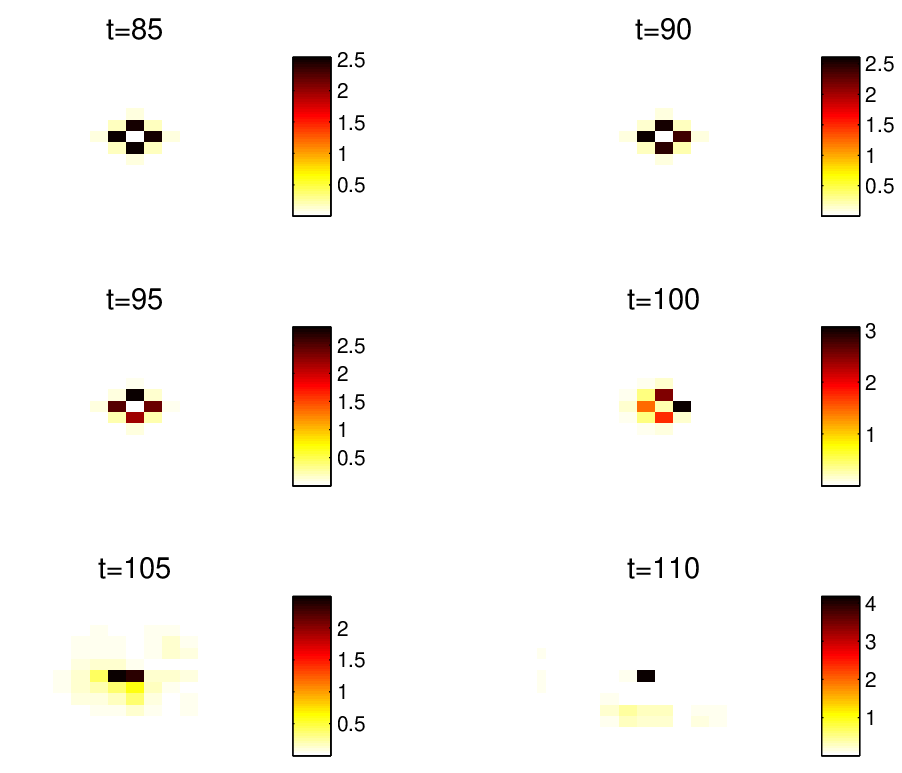} &
    \includegraphics[width=4.5cm]{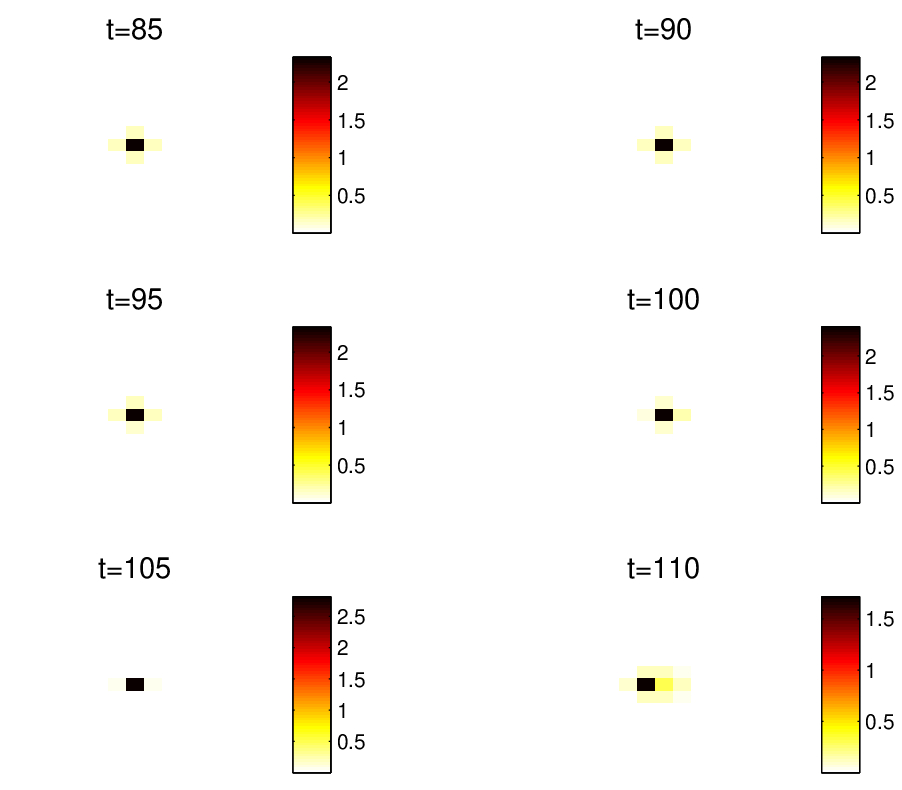} \\
\end{tabular}
\caption{Snapshots showing (a) $|U_n(t)|^2$ and (b) $|V_n(t)|^2$ for unstable interlaced vortex-solitons with $g_{12}=0.2$ and $C=0.45$.}
\label{fig9}
\end{center}
\end{figure}

The dynamics of the oscillatory unstable IVs in 2D lattices with $g_{12}=0.2$ and
$C=0.3$ is shown if Fig. \ref{fig8}. The evolution results in the transformation
of the original structure into single-peaked or multi-peaked solitons. Excited
peaks do not coincide for $U_n$ and $V_n$. The vorticity of each vortex is lost.
Fig. \ref{fig9} shows the dynamics of an oscillatorily unstable interlaced
vortex-soliton structure with $g_{12}=0.2$ and $C=0.45$. This mode evolves
spontaneously towards single-peaked solitons. The excited peaks are in the same
site in both lattices in this example.

\begin{figure}
\begin{center}
\begin{tabular}{cc}
    (a) & (b) \\
    \includegraphics[width=6.6cm,height=5cm,angle=0]{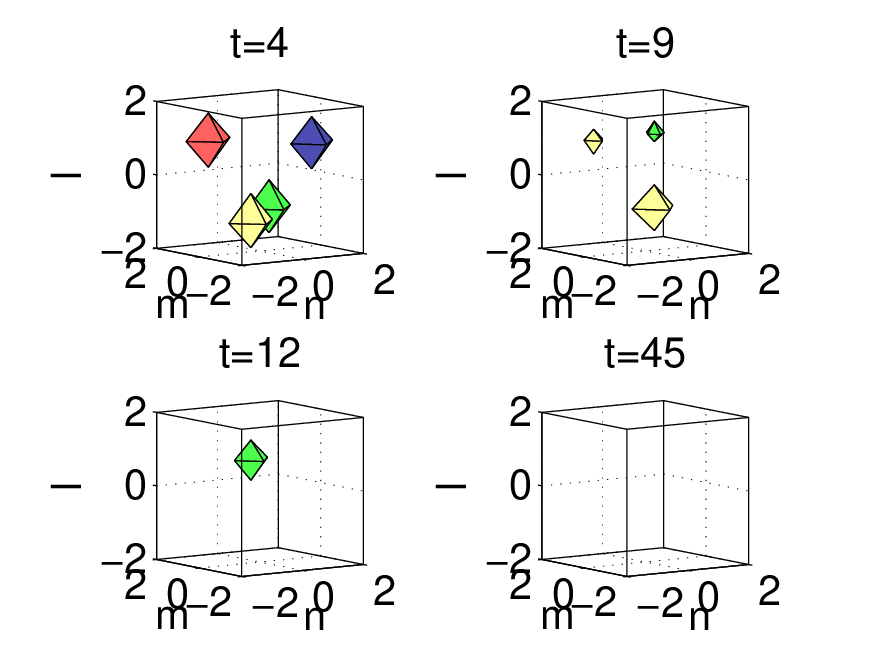} &
    \includegraphics[width=6.6cm,height=5cm,angle=0]{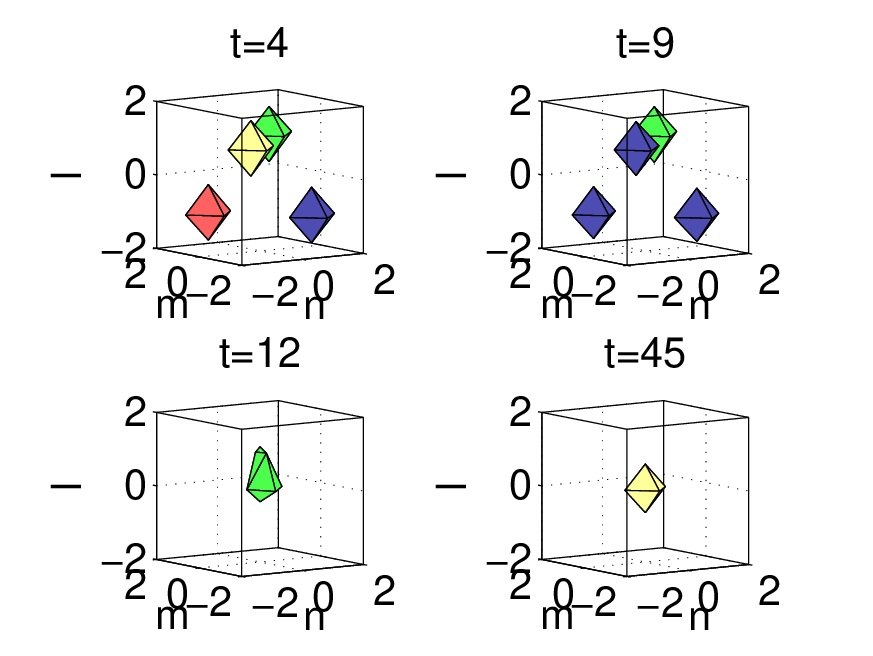} \\
\end{tabular}
\caption{Snapshots showing evolution of (a) $U_n(t)$ and (b) $V_n(t)$ for the interlaced cube
with $g_{12}=0.85$ in a grid of size $11 \times 11 \times 11$ where the coupling has been
continued to $C=0.6$. All iso-contour plots are defined as
$\mathrm{Re}(u_{n,m,l})=\mathrm{Re}(v_{n,m,l})=\pm 0.75=\mathrm{Im}(u_{n,m,l})=\mathrm{Im}(v_{n,m,l})$,
where in the figure, dark gray (blue) and gray (red) colors pertain to iso-contours of the real part of
the solutions, while the light gray (green) and very light gray (yellow) colors correspond to
the iso-contours of the imaginary part. The configuration was pertubed by a random
noise of amplitude $0.01$ in order to expedite the onset of the instability.}
\label{fig10}
\end{center}
\end{figure}

Dynamics of the interlaced cube with $g_{12} = 0.85$ is shown in Fig. \ref{fig10}.
Here the coupling is continued to $C = 0.6$. This is well past the threshold of
stability for this value of $g_{12}$, and takes the configuration into the region
of both exponential and oscillatory instabilites. It is observed that when a
peturbation
of magnitude $0.01$ is applied, only a single site survives for long times (in this
case for $V_n$).

Although these are prototypical results of the dynamical evolution, which
we have generically observed to lead to less elaborate (and often purely
single-peaked) structures in this setting, it should be stressed that the
specific details of the unstable dynamical evolution of each structure
depend considerably on the values of the parameters, as well as partially
on the type/strength of the perturbation.

\section{Conclusions and Future Challenges}

In the present work, we have illustrated the possibility to
successfully interlace structures which are stable in each
one of the components (either simple ones, such as single
site solitary waves, or more elaborate ones, such as bound states
and vortices) in order to produce stable multi-component interlaced
solitons/vortices. We have continued the resulting structures
from the anti-continuum limit of no inter-site coupling to finite
coupling and illustrated the intervals of stability, as well as the
ones of both exponential and oscillatory (Hopf) instabilities.
We have given detailed two-parameter diagrams of the stable ranges
of the solutions as a function of the inter-site and inter-component
couplings. These revealed that the linear stability of the
interlaced structures necessitates
sufficiently weak coupling (typically no larger than 0.4,
with the relevant range decreasing as the inter-component
interaction is increased) and sufficiently weak inter-component
interaction (i.e., $g_{12} < 1$). Finally, we examined the
dynamical evolution of the instability of such interlaced
structures, which typically resulted in the destruction
of the waveforms, in favor of simpler, more stable dynamical
patterns.

Nevertheless, there is still a number of important open questions
for future consideration. For instance, it would be particularly
interesting to examine whether it would be possible for the
inter-component coupling to actually stabilize structures that
are dynamically unstable in the single-component setting.
Also, it would be useful to possess a systematic classification
of the solutions (interlaced and non-interlaced ones) available
in the multi-component system setting, similarly to the
one-component classifications of \cite{pkf1,pkf2}.
Such efforts are currently underway and will be reported
in future publications.

\end{document}